\documentclass[prl,superscriptaddress,showpacs,twocolumn]{revtex4}
\usepackage{graphicx}
\usepackage{graphicx,amsfonts}
\usepackage{epsfig,amsmath}
\usepackage{verbatim}

\begin{document}

\newcommand{\atanh}
{\operatorname{atanh}}
\newcommand{\ArcTan}
{\operatorname{ArcTan}}
\newcommand{\ArcCoth}
{\operatorname{ArcCoth}}
\newcommand{\Erf}
{\operatorname{Erf}}
\newcommand{\Erfi}
{\operatorname{Erfi}}
\newcommand{\Ei}
{\operatorname{Ei}}

\title{Statistics of the Number of Zero Crossings : from Random
  Polynomials to Diffusion Equation.}

\author{Gr{\'e}gory Schehr}
\affiliation{Laboratoire de Physique Th\'eorique (UMR du
  CNRS 8627), Universit\'e de Paris-Sud, 91405 Orsay Cedex,
  France}
\author{Satya N. Majumdar}
\affiliation{Laboratoire de Physique Th\'eorique et et Mod\`eles
  Statistiques, Universit\'e Paris-Sud, B\^at. 100, 91405 Orsay Cedex,
France}

\date{\today}

\begin{abstract}
We consider a class of real random polynomials, indexed by an integer $d$, of
large degree $n$ and focus 
on the number of real roots of such random polynomials. The
probability that such polynomials have no real root in the interval
$[0,1]$ decays as a
power law $n^{-\theta(d)}$ where $\theta(d)>0$ is the exponent associated to
the decay of the persistence probability for the diffusion equation with
random initial  
conditions in space dimension $d$. For $n$
even, the probability that such polynomials have no root on the full real
axis decays
as $n^{-2(\theta(d) + \theta(2))}$. For $d=1$, this
connection allows for a physical realization of 
real random polynomials. We further show that the probability that such
polynomials have    
exactly $k$ real roots in $[0,1]$ has an unusual 
scaling form given by
$n^{-\tilde \varphi(k/\log n)}$ where $\tilde \varphi(x)$ is a universal large
deviation 
function.  
\end{abstract}
\pacs{02.50.-r, 05.40.-a,05.70.Ln, 82.40.Bj}

\maketitle

Persistence properties and related first passage problems 
have been the subject of 
intense activities, both theoretically \cite{satya_review} and experimentally
\cite{persist_exp, persist_diffusion_exp}
these last few 
years. The persistence probability $p(t)$ for a time 
dependent stochastic process of zero mean,   
is defined as the probability that it has not changed sign up to 
time $t$. In many physical situations, $p(t)$ was found to decay at large
time as a power law $p(t) \propto t^{-\theta}$. Surprisingly, the persistence
exponent $\theta$ was found to be highly non trivial even for simple 
systems. One example is the diffusion equation in space dimension $d$, 
$\partial_t \phi(x,t) = \nabla^2 \phi(x,t)$
with ``white noise" initial conditions $\langle\phi(x,0) \phi(x',0)\rangle =
\delta^d (x-x')$. The solution is characterized by a single growing length
scale $\ell(t) \propto t^{1/z}$, with $z=2$. 
For a system of linear size $L$, the persistence $p_0(t,L)$
is simply the probability that $\phi(x,t)$ does not change sign up to time
$t$. It was found \cite{persist_diffusion} that for $t\gg 1$, $p_0(t,L)$ has a
finite size scaling form
\begin{equation}
p_0(t,L) \propto L^{-z\theta(d)} h (L^z/t)
\label{fss1}
\end{equation}
with $h(u) \sim c^{\rm st}$, a constant independent of $L$ and $t$, for $u \ll
1$ and $h(u) \propto u^{\theta(d)}$ for $u \gg 1$, where $\theta(d)$ is a non
trivial exponent, {\it e.g.} $\theta(1) = 0.1207, \theta(2) =
0.1875$. The scaling form of $p_0(t,L)$ indicates that $p_0(t,L) \propto
t^{-\theta(d)}$ for large $t$ in an infinite system. Alternatively, 
$\theta(d)$ can be
extracted by measuring $p_0(t,L) \propto L^{-z\theta(d)}$ for very large
time $t \gg L^z$. Remarkably, the persistence for $d=1$ was observed in
experiments on 
magnetization of spin polarized Xe gas and the exponent $\theta_{\exp}(1)
\simeq 0.12$ was measured \cite{persist_diffusion_exp}, in good agreement with
analytical approximations and numerical simulations \cite{persist_diffusion,
  newman_diffusion}.

Another apparently unrelated problem concerns the roots of random polynomials
({\it i.e.} polynomials with random coefficients), 
which have attracted renewed interest over the last few years
\cite{edelman,rnd_poly_books} in the context of probability and number theory. A
recent work \cite{castin_complex_exp} proposed a physical realization of the
complex roots of Weyl complex polynomials in a system of a rotating
quasi-ideal atomic Bose gas. Here we focus instead on the real roots of
a class of real random polynomials indexed by an integer $d$     
\begin{eqnarray}
f_n (x) = a_0 + \sum_{i=1}^{n-1} a_i\, i^{(d-2)/4} x^i \label{real_poly_d}
\end{eqnarray}
Here $a_i$'s are real Gaussian random variables of zero mean and with
correlations $\langle a_i a_j \rangle = \delta_{ij}$. We will see
below that, for $d=2$, where $f_n(x)$ reduce to the famous Kac polynomials
\cite{kac_1}, the statistics of the real roots of $f_n(x)$ is
identical in the $4$ subintervals $]-\infty, -1[, [-1,0], [0,1]$ and
$]1,+\infty[$. However, for $d \neq 2$, the statistics 
of real roots of $f_n(x)$ depend on $d$ in the two inner intervals 
$[-1,0]$ and $[0,1]$,
while it is identical to the case $d=2$ in the two outer
ones. In this letter we will focus primarily on the interval $[0,1]$ and 
ask : what is
the probability $P_0(1,n)$ that $f_n(x)$ has no real root in $[0,1]$ ?
Recently, it was 
found, for $d=2$, that  $P_0(1,n) \propto n^{-\zeta(2)}$ for large
$n$ where the exponent $\zeta(2)\approx 0.19(1)$ was computed numerically
\cite{dembo}. In addition, for the special case $d=2$, the
authors of Ref. \cite{dembo}
showed that $P_0(1,n)$ is related to the probability of no zero crossing
of a Gaussian 
stationary process (GSP) with correlator ${\rm sech}(|T|/2)$. 

The purpose of this Letter is to provide a link between the persistence
of the diffusion equation and the probability $P_0(1,n)$ that $f_n(x)$ has no
real root in $[0,1]$. For arbitrary dimension $d$, we show that $P_0(1,n) \propto
n^{-\zeta(d)}$ with $\zeta(d) = \theta(d)$. Given that $\theta(1)$ was measured
experimentally \cite{persist_diffusion_exp}, this demonstrates an experimental
realization of real random polynomials. The
connection between these two 
problems in arbitrary $d$ is achieved by showing that
both problems can be mapped to the same GSP. 
In addition, we compute the probability that 
a "smooth" GSP, such as the one that appears in the context of
diffusion equation, 
crosses zero exactly $k$ times up to time $T$. Translated into
the language of random polynomials,  
our analysis shows that the probability $P_k(1,n)$ that
$f_n(x)$ has exactly $k$ real roots in the interval $[0,1]$ has a
rather unusual scaling form 
(for large $k$, large $n$, but keeping the ratio $k/\log n$ fixed)
\begin{eqnarray}
P_k(1,n) \propto n^{-\tilde\varphi\left(\tfrac{k}{\log n} \right)}
\label{rpscaling1}
\end{eqnarray}           
where $\tilde \varphi(x)$ is a large deviation function, with $\tilde
\varphi(0) = 
\zeta(d)$. Besides, our numerical analysis suggests that $\tilde \varphi(x)$ is
universal in the sense that it is independent of the 
distribution of $a_i$ provided $\langle a_i^2 \rangle$ is finite.

To study the persistence probability $p_0(t,L)$ of the diffusion
equation, it is customary to study the normalized
process $X(t) = \phi(x,t)/\langle \phi(x,t)^2\rangle^{1/2}$
\cite{satya_review}. Its autocorrelation 
function $a(t,t') = \langle X(t) X(t') \rangle$
is given, in the limit $t,t' \ll L^2$ by $a(t,t') = [4
  tt'/(t+t')^2]^{d/4}$. In terms of  
logarithmic time variable $T = \log t$, $X(T)$ is a GSP with correlator 
$a(T,T') \equiv a(T-T') = [{\rm sech}(|T-T'|/2)]^{d/2}$, which decays
exponentially for large $|T-T'|$. Thus the persistence probability $p_0(t,L)$,
for $t\ll L^2$, reduces to the computation of the probability ${\cal P}_0(T)$
of no zero 
crossing of $X(T)$ in the interval $[0,T]$. It is well known \cite{slepian}
that if $a(T) < 1/T$ at large $T$ then ${\cal P}_0(T)\sim \exp[-\theta T]$
decays exponentially for large $T$ where the decay constant $\theta$ depends
on the full stationary correlator $a(T)$. Reverting back to 
the original time $t=e^T$, one
finds $p_0(t,L) \sim 
t^{-\theta(d)}$, for $t \ll L^{2}$. In the opposite limit $t \gg L^2$, one has
$p_0(t,L) \to A_L$, a constant which depends on $L$. These two
limiting behaviors 
of $p_0(t,L)$ can be combined into a single finite size scaling form 
in Eq.~(\ref{fss1}) 
where $\theta(d)$ is the decay constant associated with the no zero crossing
probability of the   
GSP with correlator $a(T)=[{\rm sech}(|T|/2)]^{d/2}$
\cite{persist_diffusion}.   

The mapping of real random polynomials $f_n(x)$
in Eq.~(\ref{real_poly_d}) to a GSP is more subtle. 
We first observe that for large $n$ the real roots of $f_n(x)$ are
concentrated around $x = \pm 1$. To show this, we have
generalized the Kac's method \cite{kac_1} to compute 
the mean density of real roots $\rho_n(x)$ of $f_n(x)$ 
(\ref{real_poly_d}). We do not give the details of the computation and simply
quote the results here. 
We find that $\rho_n(\pm 1)$ diverges in the large $n$ limit as       
$\rho_n(\pm 1) \sim 2n \sqrt{d/(d+4)}/(\pi(d+2))$ and away from these
singularities $\rho_{\infty}(x \neq \pm 1)$ is given, for $|x| < 1 $, by
\begin{equation}
\rho_{\infty}(x) = \frac{({
    {\rm Li}_{-1-d/2}(x^2)(1+{\rm Li}_{1-d/2}(x^2)) - 
    {\rm Li}^2_{-d/2}(x^2)  })^{\tfrac{1}{2}}} 
{\pi|x|(1+{\rm Li}_{1-d/2}(x^2))} \label{mean_density}
\end{equation} 
where ${\rm Li}_n(z) = \sum_{i=1}^\infty z^i/i^n$ is the polylogarithm
function, yielding back $\rho_{\infty}(x \neq \pm 1) = (\pi|1-x^2|)^{-1}$ for
$d=2$. In particular, one has $\rho_{\infty}(0) = 1/\pi$ for all 
$d$, and $\rho_{\infty}(x) \sim (d/2)^{\tfrac{1}{2}}(2\pi((1-x)))^{-1}$
for $x \to 1^-$. For $|x| > 1$, $\rho_{\infty}(x)=1/[\pi(x^2-1)]$ for all
$d$ \cite{das}.
\begin{figure}
\includegraphics[angle=0,width=1\linewidth]{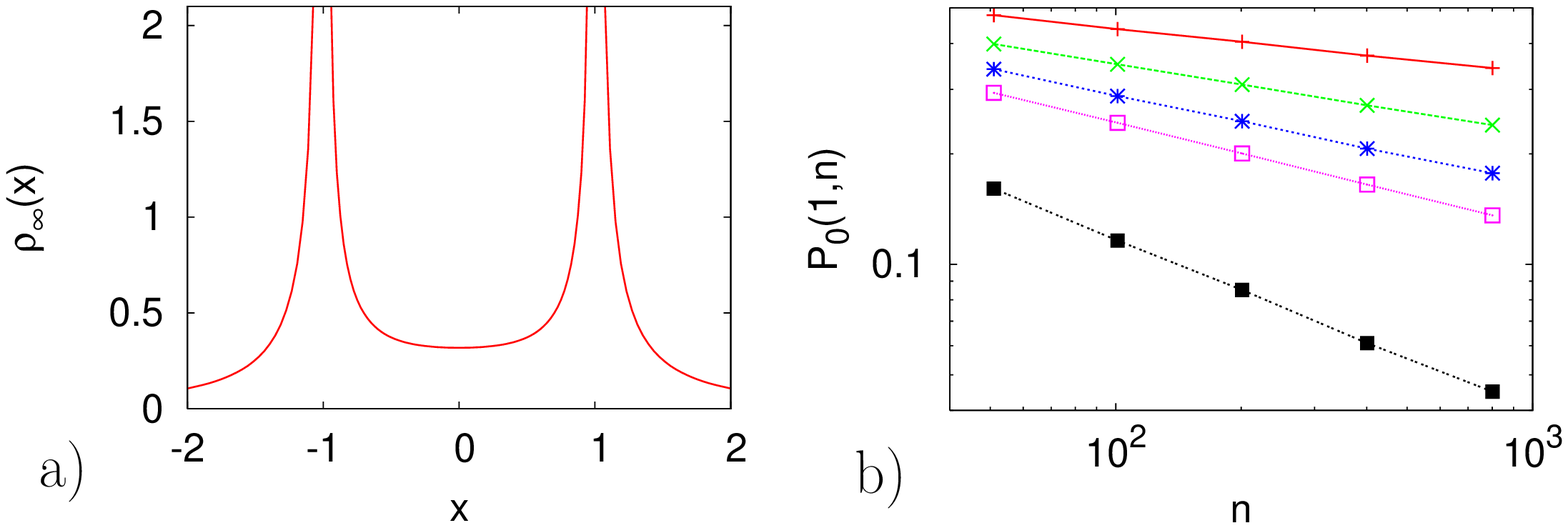}
\caption{{\bf a) }: $\rho_{\infty}(x)$ given analytically in
  Eq. (\ref{mean_density}) and below 
as a function of $x$ for $d=1$. The divergences for $x = \pm 1$ indicate that
the real roots concentrate around $x = \pm 1$ for large $n$. {\bf b)} 
: Plot of $P_0(1,n)$ as a function of $n$ for different values of
$d = 1,2,3,4,10$. The lines are guide to the eyes. The measured exponents
are given by $0.12, 0.18, 0.23, 0.27, 0.46$ for $d = 1,2,3,4,10$
respectively, in full agreement with the numerical values 
for $\theta(d)$ obtained in Ref. \cite{persist_diffusion, newman_diffusion}.} 
\label{fig1} 
\end{figure}
In Fig. \ref{fig1} a), we plot
$\rho_{\infty}(x)$ for 
$d=1$ where the divergence at $x = \pm 1$ indicates that
the real roots 
concentrate around $x = \pm 1$ for large $n$. 

The random polynomial $f_n(x)$ being a Gaussian process, it
is completely 
determined by its two-point correlator $C_n(x,x') = \langle
f_n(x) f_n(x')\rangle$. Using the fact that $a_i$'s are uncorrelated,
it is easy to see that
\begin{equation}
C_n(x,x')= 1+\sum_{i=1}^{n-1} i^{(d-2)/2}\, (xx')^i .
\label{corr1}
\end{equation} 
It is useful to introduce the normalized
correlator ${\cal C}_n 
(x,x') =  C_n(x,x')/{ C_n(x,x)^{\tfrac{1}{2}}
 C_n(x',x')^{\tfrac{1}{2}}}$. From the analysis of ${\cal C}_n 
(x,x')$ in the large $n$ limit, one can show that, asymptotically, $f_n(x)$
takes 
independent values in 
the $4$ subintervals $]-\infty, -1[, [-1,0], [0,1]$ and $]1,+\infty[$.
For $d=2$, ${\cal C}_n (x,x')$ has a special symmetry, namely
${\cal C}_n (x,x')={\cal C}_n
(-x,-x')={\cal C}_n
(1/x,1/x')$, meaning that the normalized Gaussian processes in
these $4$ subintervals are independent and isomorphic. Denoting
$P_0(1,n)$ as the probability of no zero crossing in 
$[0,1]$, it follows that the probability $q_0(n)$ that there are no
real roots for $d=2$ is precisely equal to $[P_0(1,n)]^4$ for large
$n$. For $d \neq 2$, one can show that the behavior of ${\cal C}_n 
(x,x')$ depends on $d$ in the two inner intervals (see below), while it
behaves like for $d=2$ in the two outer ones \cite{us_inprep}.

We now focus on the interval $[0,1]$ and 
to make a
precise connnection with the persistence probability $p_0(t,L)$ 
defined in the context of the diffusion equation we define $P_0(x,n)$,
for $x \leq 1$, as the probability that $f_n(x)$ has no real root in the
interval $[0,x]$. 
We next reparametrize the polynomial with
a change of variable, $x = 1-1/t$.  One 
finds that the relevant scaling limit of ${\cal C}_n(t,t')$ is
obtained for $t,t',n \to \infty$ keeping $\tilde t = t/n$ and $\tilde t' =
t'/n$ fixed (see also Ref. \cite{fyodorov}). In that scaling limit one finds
that ${\cal C}_n(t,t') \to {\cal 
  C}(\tilde t, \tilde t')$ with the asymptotic behaviors  
\begin{eqnarray} \label{correlator_asympt}
{\cal C}(\tilde t,\tilde t') \sim 
\begin{cases}
\left(4 \frac{{\tilde t \tilde t'}}{(\tilde t+\tilde
  t')^2}\right)^{\tfrac{d}{4}} \,,& \tilde t, \tilde 
t' \ll 1 \\ 
1 \,, & \tilde t,\tilde t' \gg 1 
\end{cases}
\end{eqnarray}
For
$\tilde t,\tilde t' \ll 1$, this correlator is exactly the same as the one
found for diffusion, ${\cal C}(\tilde t, \tilde t') = a(t,t')$ in this regime. 
Since a Gaussian process is completely characterized by its two point
correlator, 
we conclude that the diffusion process and the random polynomial 
are essentially the same Gaussian process and hence have the same
zero crossing properties.
In the opposite limit, $\tilde t,\tilde t' \gg 1$ 
the fact that ${\cal C}(\tilde t,\tilde t') \to 1$ suggests that $P_0(x,n)$
goes to a constant when $x \to 1$. Therefore, in complete analogy 
with Eq. (\ref{fss1}) 
we propose the scaling form for random polynomials 
\begin{eqnarray}
  P_0(x,n) \propto n^{-\theta(d)} \tilde h(n (1-x)) \label{persist_poly}
\end{eqnarray}   
with $\tilde h(u) \sim c^{st}$ for $u \ll 1$ and $\tilde h(u) \sim
u^{\theta(d)}$ for $u \gg 1$, where 
$\theta(d)$ is the persitence exponent associated to the diffusion 
equation in dimension $d$. Note that $n$ here plays the role of
$L^z$ in diffusion problem while the variable $1-x$ is the 
analogue of the inverse time $1/t$. 
This scaling form (\ref{persist_poly}), which we
verify numerically, thus 
establishes a direct link between the real roots of real random polynomials
and the diffusion equation with random initial conditions. It also follows
from our earlier discussion on the $4$ subintervals that the
probability that $f_n(x)$ has no real root decays like   
$n^{-b(d)}$ with $b(d) = 2 (\theta(d)+\theta(2))$ where $2\theta(d)$
is contributed by the 2 inner intervals and $2\theta(2)$ 
by the two outer intervals.   

\begin{figure}
\includegraphics[angle=0,width=1\linewidth]{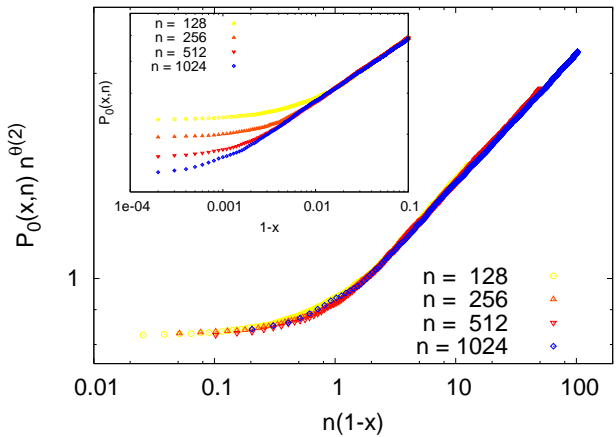}
\caption{Plot of $n^{\theta(2)} P_0(x,n)$, with $\theta(2) = 0.1875$ as a
  function of $(1-x)n$ for different degrees $n = 64, 128, 256, 512,
  1024$. {\bf Inset} : $P_0(x,n)$ as a function of $1-x$ for  
  different degrees $n$.}
\label{fig2} 
\end{figure}

We have verified  this scaling form (\ref{persist_poly}) numerically
by computing 
the number of real roots 
in the interval $[0,x]$ of random polynomials such as in
Eq. (\ref{real_poly_d}) for different degrees $n$. In each case, the
probability distribution $P_0(x,n)$ is obtained by averaging over 
$10^4$ realizations of the random variables $a_i$'s, drawn independently from  
a Gaussian
distribution of unit variance. The plot shown on
Fig. \ref{fig2}, for $d=2$, shows a good agreement with the scaling in
Eq. (\ref{persist_poly}) with $\theta(2)=0.1875$, in agreement with the
numerical value reported in Ref.~\cite{dembo}.     
We have also calculated the exponent 
$\theta(d)$ in Eq. (\ref{persist_poly}),  for various dimensions, by
measuring $P_0(1,n)$ for different 
values of $d$. The results are shown in Fig. \ref{fig1} b) where, according
to our 
prediction in Eq. (\ref{persist_poly}) $P_0(1,n)\sim n^{-\theta(d)}$ for large $n$.
The value of
$\theta(d)$ obtained this way is fully compatible with previous
numerical estimates \cite{persist_diffusion, newman_diffusion} of
$\theta(d)$ for the diffusion equation.  

\begin{figure}[h]
\includegraphics[angle=0,width=1\linewidth]{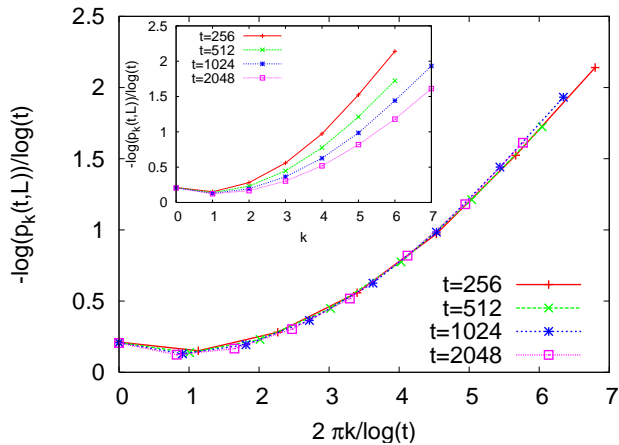}
\caption{$(-\log{p_k(t,L)})/\log{t}$ for the diffusion equation with random
  initial conditions as a function of $k/\mu \log{t}$ with
  $\mu =1/2\pi$ for different times $t=256, 512, 1024, 2048$. {\bf Inset }: $(-\log{p_k(t,L)})/\log{t}$ as a function of $k$ for different times. }  
\label{fig4}
\end{figure}

We now generalize our analysis 
and consider the probability $p_k(t,L)$ that the diffusing field $\phi(x,t)$
crosses zero exactly $k$ times up to time $t$ (similarly, one considers
the probability $P_k(x,n)$, $x<1$, that such polynomials
(\ref{real_poly_d}) have 
exactly $k$ 
real roots \cite{dembo} in the interval $[0,x]$, see below). Let
us first consider the regime
$1 \ll t \ll L^z$. In this regime, $p_k(t,L)$
is 
given by the probability ${\cal P}_k(T)$ that $X(T)$ crosses zero exactly $k$
times where $X(T)$ is a GSP with correlations $a(|T-T'|) =
[{\rm sech}(|T-T'|)]^{d/2}$, where $T=\log t$. 
Since, $a(T) = 1 - \tfrac{d}{16} T^2 + o(T^2)$ for small $T$,
our GSP is a smooth process with a finite density of zero crossings 
given by the
Rice's formula $\mu = (-{a''(0)})^{\tfrac{1}{2}}/\pi$
\cite{rice_formula}. We propose the following scaling form for
large $T$ and large~$k$
\begin{eqnarray}
\log{{\cal P}_k(T)} = -T \varphi\left(\frac{k}{\mu T}
\right). 
\label{scaling}
\end{eqnarray} 
To understand the origin of this scaling form, let us
consider the generating function $\hat {\cal P}(p,T) = \sum_{k=0}^\infty p^k
{\cal P}(k,T)$ \cite{satya_partial}.
It turns out that, $\hat {\cal P}(p,T) \sim \exp(-\hat \theta(p) T)$,
where for a smooth GSP $\hat \theta(p)$ depends
continuously on $p$ : this was shown exactly for the random
acceleration process and approximately using the independent interval
approximation for arbitrary smooth GSP -- and checked numerically for the
diffusion equation 
with random initial conditions \cite{satya_partial}. If the
scaling in Eq. (\ref{scaling}) holds, one gets by steepest
descent method valid for large $T$, $\hat \theta(p) =
{\rm Min}_{x > 0} 
[\mu x \log p - \varphi(x)]$. Inverting the Legendre transform we get 
\begin{eqnarray}
\varphi(x) = {\rm Max}_{0 \leq p \leq 2}[ \mu x \log p + \hat \theta(p)]
\label{legendre} 
\end{eqnarray}
Notice that although $\hat \theta(p)$ is a priori defined on the interval
$[0,1]$, the computation of $\varphi(x)$ involves an
analytical continuation of $\hat \theta(p)$ on $[0,2]$. Going back to real
time $t$, Eq. (\ref{scaling}) then yields a rather unusual scaling form valid in the
limit $1\ll t \ll L^2$ 
\begin{eqnarray}
p_k(t,L) \sim t^{-\varphi\left(\tfrac{k}{\mu \log t}\right)}
\label{peculiar_scaling} 
\end{eqnarray}
For $k$ close to $\mu \log t$, one expects $p_k(t,L)$ to behave locally as a
Gaussian and $\varphi(x)$ is thus quadratic around $x=1$. Away from the
minimum, 
we have not been able to obtain $\varphi(x)$ analytically. 
We have thus tested the scaling form
(\ref{peculiar_scaling}) numerically. We used a space time discretized
diffusion equation 
\begin{eqnarray}
\phi_i(t+1) = \phi_i(t) + a \sum_{j}[\phi_j(t) - \phi_i(t)]
\end{eqnarray} 
where $j$ runs over the nearest neighbours of $i$ on a $d$-dimensional square
lattice. A
stability analysis shows that the solution is unstable for $a > a_c=1/2d$ and we
chose $a=a_c/2$ that provided the quickest onset of the asymptotic
behavior. The initial values $\phi_i(0)$'s were
chosen independently from a Gaussian distribution of zero mean and unit
variance.  
Fig. \ref{fig4} shows the results for $d=2$, {\it i.e.} the case of
Kac's polynomials (\ref{real_poly_d}).
We computed $p_k(t,L)$ for different times
$t$ for a system of linear size $L = 256$ and averaging 
over $100$ different realizations of the random
initial condition. In the inset of Fig. \ref{fig4}, we plot 
$(-\log{p_k(t,L)})/\log{t}$ as a 
function $k$ where the different curves correspond to different times $t$. 
In Fig. \ref{fig4} these curves for different time $t$ fall on the
same master curve when plotted as a function of  
the rescaled variable $k/\mu \log{t}$ with $\mu =1/2\pi$ in this case,
confirming the validity of the scaling form in
Eq. (\ref{peculiar_scaling}). We also checked
that a different 
choice of the 
distribution of $\phi_i(0)$, such as $\phi_i(0) = \pm 1$ or rectangular,
gave, within the error bars, the same function $\varphi(x)$ thus indicating the
universality of this large deviation function.  

In the opposite limit $t\gg L^z$, one simply replaces $t$ in
(\ref{peculiar_scaling}) 
by $L^z$. Translating into random polynomials, this regime corresponds to
 $(1-x) \ll n^{-1}$ since one just replaces $t$ by $1/(1-x)$ and 
$L^z$ by the degree $n$ as discussed before. Thus, 
in this regime, we arrive 
at the announced scaling
form for $q_k(n)$ in Eq. (\ref{rpscaling1}).  
For the special case of Kac's polynomials ($d=2$), this scaling form, 
in the neighbourhood of $k=\log n/2\pi$, is consistent with the rigourous
result \cite{maslova} that in this neighbourhood $q_k(n)$
is a Gaussian with mean $\log n/2\pi$ and variance 
$V_n \sim
\tfrac{1}{\pi}(1-\tfrac{2}{\pi})\log n $ in the
large $n$ limit. 

In fact, the scaling in Eq. (\ref{peculiar_scaling}) holds more generally for
any smooth GSP. To illustrate this, we
considered the random 
acceleration 
process $d^2 x(t)/dt^2 = \eta(t)$ where $\eta(t)$ is a white noise, and for
which $\mu = \sqrt{3}/(2\pi)$. For this
particular smooth GSP $\hat \theta(p)$ has been computed exactly
\cite{burkhardt}, yielding $\hat \theta(p) =
\tfrac{1}{4}(1-\tfrac{6}{\pi}\sin^{-1}(\tfrac{p}{2}))$. By performing the
Legendre transform (\ref{legendre}) one obtains the asymptotic behaviors as
\begin{eqnarray}\label{large_dev_rnd_acc}     
\varphi(x) \sim
\begin{cases}
\tfrac{1}{4} + \tfrac{\sqrt{3}}{2\pi} x \log x \quad, \quad x \to 0 \\
\tfrac{3\sqrt{3}}{16 \pi} (x-1)^2 \quad, \quad x \to 1 \\
\tfrac{\sqrt{3}}{2\pi} x \log{2} \quad, \quad x \to \infty
\end{cases}
\end{eqnarray}
which gives back the exact result $\varphi(0) = 1/4$ \cite{sinai_burckhardt}. 

To conclude, we have established in this Letter a connection between the
persistence 
probability $p_0(t,L)$ for the diffusion equation in dimension $d$ and the
probability $P_0(x,n)$ that generalized Kac's 
real random polynomials, indexed by $d$ as in
Eq. (\ref{real_poly_d}), have no real roots in the interval $[0,x]$, with $x <
1$. This connection is useful in predicting new results for random
polynomials, such as the unusual scaling form (\ref{rpscaling1}) for the
probability of having $k$ real roots in $[0,1]$. Besides, we hope that
this connection 
may also shed some light in calculating the exponents $\theta(d)$ exactly
which still remains a challenge.

\acknowledgments
S.N.M thanks J.~Unterberger for pointing out Ref.~\cite{dembo}.


\begin{thebibliography}{32}

\bibitem{satya_review}
S.N. Majumdar, Curr. Sci., {\bf 77}, 370 (1999). 


\bibitem{persist_exp}
W.Y. Tam {\it et al}, Phys. Rev. Lett. {\bf 78},
1588 (1997);  
D. B. Dougherty {\it et al}, Phys. Rev. Lett. {\bf 89}, 136102 (2002).   

\bibitem{persist_diffusion_exp}
G.P. Wong {\it et al}, Phys. Rev. Lett. {\bf 86}, 4156 (2001).


\bibitem{persist_diffusion}
S.N. Majumdar {\it et al}, Phys. Rev. Lett. {\bf 77},
2867 (1996); B. Derrida, V. Hakim and R. Zeitak, ibid. 2871.



\bibitem{newman_diffusion}
T. J. Newman and W. Loinaz, Phys. Rev. Lett. {\bf 86}, 2712 (2001). 

\bibitem{edelman}
A. Edelman and E. Kostlan, Bull. Amer. Math. Soc. {\bf 32}, 1 (1995).

\bibitem{rnd_poly_books}
K. Farahmand, in {\it Topics in random polynomials}, Pitman research notes in
mathematics series {\bf 393}, (Longman, Harlow) (1998).  


\bibitem{castin_complex_exp}
Y. Castin {\it et al}, Phys. Rev. Lett. {\bf 96}, 040405 (2006). 

\bibitem{kac_1}
M.~Kac, Bull. Amer. Math. Soc. {\bf 49}, 314 (1943); Erratum:
Bull. Amer. Math. Soc. {\bf 49}, 938 (1943).   


\bibitem{dembo}
A.~Dembo {\it et al}, J. Amer. Math. Soc. {\bf 15},
857 (2002). 


\bibitem{slepian}
D. Slepian, Bell Syst. Tech. J. {\bf 41}, 463 (1962).

\bibitem{das}
M. Das, J. Indian Math. Soc. {\bf 36}, 53 (1972).


\bibitem{rice_formula}
S.O.~Rice, Bell Syst. Tech. J, {\bf 23}, 282 (1944).


\bibitem{satya_partial}
S. N. Majumdar and A. J. Bray, Phys. Rev. Lett. {\bf 81}, 2626 (1998).  

\bibitem{maslova}
N.B. Maslova, Theor. Proba. Appl. {\bf 19}, 461 (1974). 

\bibitem{us_inprep}
G. Schehr and S.~N.~Majumdar, in preparation. 


\bibitem{fyodorov}
A.P. Aldous and Y.V.Fyodorov, J. Phys. A: Math. Gen. {\bf 37}, 1231 (2004). 


\bibitem{burkhardt}
T. W. Burkhardt, Phys. Rev. E {\bf 63}, 011111 (2001). 

\bibitem{sinai_burckhardt}
Y.G. Sinai, Theor. Math. Phys. {\bf 90}, 219 (1992); T.W. Burkhardt,
J. Phys. A : Math. Gen. {\bf 26}, L1157 (1993). 


\end{thebibliography}
\end{document}